\documentclass[a4paper]{article}

\usepackage{INTERSPEECH2020}

\usepackage{hyperref}
\usepackage{multirow}
\usepackage{makecell}
\usepackage{cite}
\usepackage{bm}
\usepackage{url}
\def\R{\textbf{R}}
\def\C{\textbf{C}}
\def\D{\textbf{D}}
\def\S{\textbf{S}}
\def\X{\bm{X}}
\def\H{\bm{H}}
\def\x{\bm{x}}
\def\y{\bm{y}}
\def\h{\bm{h}}
\def\p{\bm{p}}
\def\Y{\bm{Y}}
\def\P{\bm{P}}

\title{Recognition-Synthesis Based Non-Parallel Voice Conversion with \\Adversarial Learning}
\name{Jing-Xuan Zhang, Zhen-Hua Ling, Li-Rong Dai\thanks{This work was partially funded by the National Natural Science Foundation
of China (Grants No. 61871358 and U1613211), National Key R\&D Program of China (2017YFB1002202) and the Key Science \& Technology Project of Anhui Provinces (18030901016).}}
\address{National Engineering Laboratory for Speech and Language Information Processing,\\
University of Science and Technology of China, Hefei, P.R.China}
\email{nosisi@mail.ustc.edu.cn, \{zhling, lrdai\}@ustc.edu.cn}

\begin{document}

\maketitle
\begin{abstract}
 This paper presents an adversarial learning method for recognition-synthesis based non-parallel voice conversion.
A recognizer is used to transform acoustic features into linguistic representations
while a synthesizer recovers output features from the recognizer outputs
together with the speaker identity.
By separating the speaker characteristics from the linguistic representations,
voice conversion can be achieved by replacing the speaker identity with the
target one.
In our proposed method, 
a speaker adversarial loss is adopted  in order to obtain speaker-independent linguistic
representations using the recognizer.
Furthermore, discriminators
are introduced and a generative adversarial network (GAN) loss
is used to prevent the predicted features from being over-smoothed.
For training model parameters, a strategy of pre-training on a multi-speaker dataset
 and then fine-tuning on
 the source-target speaker pair is designed.
Our method achieved
higher similarity than the baseline model that obtained the best performance
in Voice Conversion Challenge 2018.
\end{abstract}

\section{Introduction}

Voice conversion (VC) aims to modify a source utterance into an output utterance, which sounds as if it is uttered by a target speaker but keeps the linguistic contents unchanged \cite{Childers1985Voice, Childers1989Voice}. In recent years, neural networks, such as deep neural networks (DNN) \cite{Desai2009voice, Desai2010Spectral}, recurrent neural networks (RNN) \cite{Sun2015Voice, nakashika2015voice} and sequence-to-sequence (seq2seq) networks \cite{8607053, tanaka2019, zhangnonpara2020}, have been applied to build the acoustic models for voice conversion and achieved great success.

According to the characteristic of training data, VC methods can be roughly categorized into two classes, i.e. parallel VC and non-parallel VC \cite{Mohammadi2017An}. In parallel VC, an acoustic model is trained with paired source-target acoustic frames or sequences. However, it's difficult to do so in non-parallel VC due to the lack of parallel training data. Many methods have been proposed for non-parallel VC and recognition-synthesis (Rec-Syn) is one of them \cite{7900072, sun2016phonetic, miyoshi2017voice, ljliu2018wav, liu2018voice}. At the conversion stage of this method, an automatic speech recognition (ASR) model is first employed to extract linguistic-related features, e.g. phonetic posteriorgrams (PPGs) \cite{sun2016phonetic} or bottleneck features \cite{ljliu2018wav}, from the source speech. Then, a synthesis model is applied to predict the acoustic features of the target speaker.
However, without explicitly disentangling linguistic and speaker representations,
the outputs of the ASR model often contain the information of source speakers, which may harm the similarity of converted voice.
Besides, the converted voice often suffers from the over-smoothing issue \cite{bishop1994mixture} because the mean square error (MSE) criterion is usually adopted for training the synthesis model.

To overcome these limitations, an adversarial learning method for Rec-Syn based non-parallel VC is presented in this paper.
In our method, a recognizer is adopted for extracting linguistic representations
and a synthesizer is adopted for predicting the converted acoustic features.
When extracting linguistic representations,
a speaker adversarial learning loss is employed besides the phoneme recognition loss, thus the
linguistic representations are processed to be speaker-agnostic.
Also, generative adversarial network (GAN) losses \cite{goodfellow2014} are used in order to alleviate the over-smoothing effect.
The WaveNet vocoder \cite{denoord2016wavenet} is adopted for recovering the waveforms of converted voice.
For training model parameters, an external multi-speaker dataset is first adopted for pre-training.
Then, the model is adapted to the desired conversion pair by fine-tuning.

Experiments are conducted to compare our method with a Rec-Syn baseline, which achieved the best performance in Voice Conversion Challenge 2018 \cite{ljliu2018wav}. The experimental results showed that our proposed method obtained better performance, especially on the similarity of converted speech. Ablation studies were also carried out to demonstrate the effectiveness of several important components in our proposed model.

\section{Related Work}
Our method is similar to the auto-encoder (AE) based VC with speaker adversarial learning
\cite{polyak2019, chou2018multi, ocal2019, 9054734}.
Polyak \emph{et al.} \cite{polyak2019} proposed a WaveNet based AE model for VC with a speaker confusion network. Chou \emph{et al.} \cite{chou2018multi}  employed an adversarial trained AE for VC and the voice quality is further improved by another residual generator and discriminator.
In both our method and previous studies, the acoustic features are first transformed into speaker-independent representations
, which are then decoded back into acoustic features.
The main difference between our method and the AE-based VC is that our method
utilizes text supervision for building the ASR module and extracting linguistic representations explicitly at training stage.
Therefore, our method belongs to the category of Rec-Syn based VC rather than the AE-based one.

\begin{figure}
  \centering
  \includegraphics[width=1.0\linewidth]{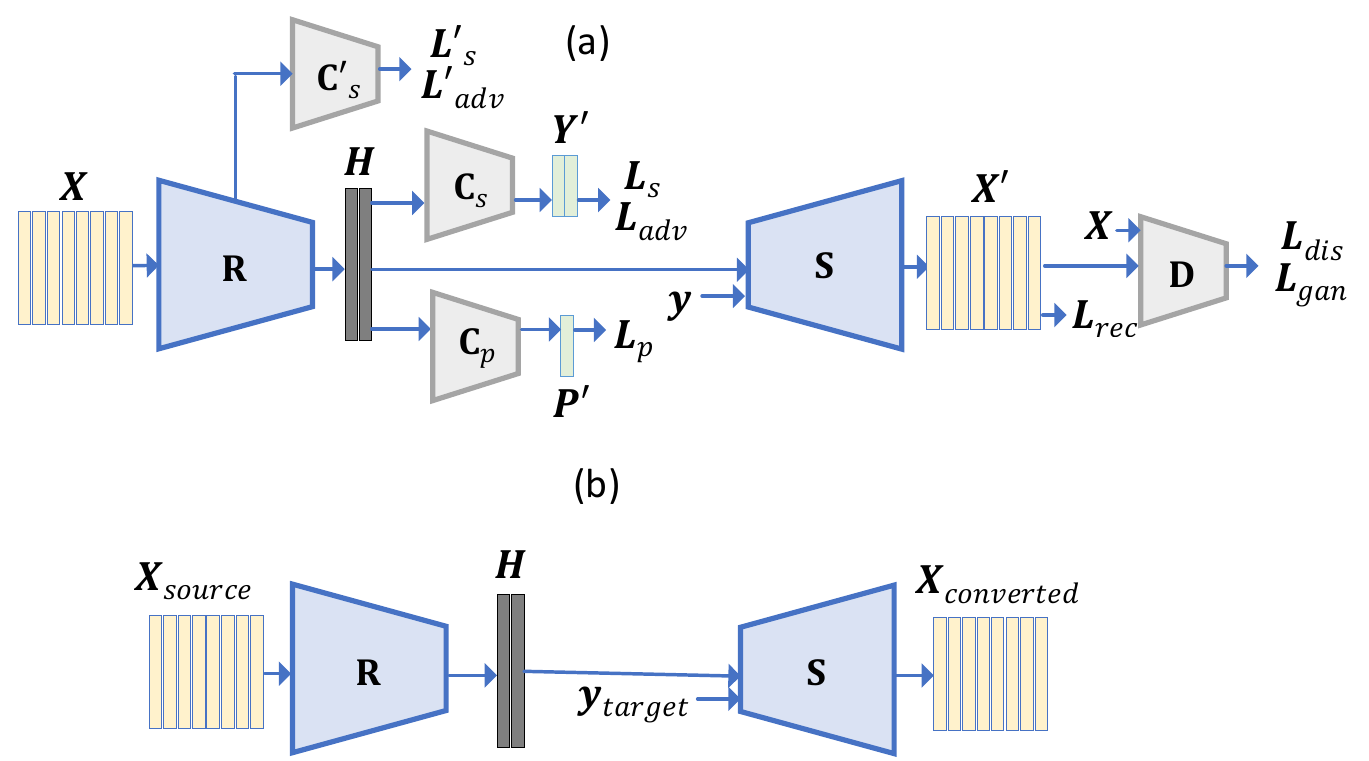}\\
  \caption{(a) The diagram of our proposed method at training stage.
  (b) The conversion process of our proposed method. $\X$, $\H$ and $\y$ represent acoustic features, linguistic representations and speaker label respectively.}\label{fig:sys}
\end{figure}

\section{Proposed Method}
\subsection{Structure overview}
Our model is consist of a recognizer $\R$ for transforming the acoustic features into linguistic representations,
a phoneme classifier $\C_p$ for phoneme label classification,
a speaker classifier $\C_s$ for eliminating speaker information, a synthesizer $\S$ for recovering acoustic features, and discriminators $\D$ for obtaining GAN losses.
\figurename~\ref{fig:sys} (a) depicts the overall structure of the proposed method at training stage.
During conversion, $\C_p$, $\C_s$ and $\D$ are discarded as shown in \figurename~\ref{fig:sys} (b). Details and training losses of these components are described in the following subsections.

\subsection{Recognition process}
Linguistic representations are extracted by the recognizer as $\H = \R(\X)$,
where $\X = [\x_1, \dots, \x_{N_x}]$ and $N_x$ are acoustic features and its frame number respectively.  $\H = [\h_1, \dots, \h_{N_h}]$ and $N_h$ are linguistic representations and its frame number respectively.
The recognizer is built with two-layer bi-directional LSTM interleaved with strided CNN. It decreases the sampling rate of input sequences by 4 times thus we have $N_h = N_x /4$.

With inputs of linguistic representations, the phoneme classifier predicts the sequence of phoneme labels
as $\P'= \C_p(\H)$,
where $\P' = [\p'_1, \dots, \p'_{N_p}]$ and $N_p$ is the length of phoneme sequence.
$\C_p$ is one-layer LSTM equipped with attention module
\cite{chorowski2015attention} and auto-regressive connection.
A cross-entropy loss is used as
\begin{equation}
L_{p} =  \frac{1}{N_p} \Sigma_{n=1}^{N_p}{\text{CE}(\p_n, \p'_n)}.
\end{equation}

The speaker classifier tries to infer the speaker identity from linguistic representations as $\Y'=\C_s(\H)$ frame by frame, where each frame in $\Y'=[\y'_1, \dots, \y'_{N_h}]$ is the probability distribution of the predicted speaker.
It is built with a 3-layer CNN.
A cross entropy loss of speaker classification is used for $\C_s$ as
\begin{equation}
L_{s} = \frac{1}{N_h} \Sigma_{n=1}^{N_h}{\text{CE}(\y, \y'_n)},
\end{equation}
where $\y$ represents the ground-truth speaker label encoded as one-hot vector.
Meanwhile, the recognizer is trained adversarially to make $\H$ speaker-invariant. As suggested in previous studies of learning disentangled representations \cite{zhou2019talking}, a speaker adversarial loss is applied to the recognizer as
\begin{equation}
\label{eq:adv}
L_{adv} = \frac{1}{N_h} \Sigma_{n=1}^{N_h}{\text{MSE}(\frac{1}{|\y|}, \y'_n)},
\end{equation}
where $|\y|$ represents the number of speakers in the training dataset.
Therefore, the loss penalizes the distance between prior and predicted
distribution of speaker probabilities.
To strengthen the adversarial training, a secondary speaker classifier $\textbf{C}'_s$ is also applied to the outputs of the first LSTM layer in $\R$. And it's also trained with a classification loss $L_s'$ and passes an adversarial loss $L_{adv'}$.

As indicated by Ocal \emph{et al.} \cite{ocal2019}, the error rate of the optimal speaker classifier relates to an upper bound of mutual information $I(\y;\H)$.
 In order to approximate the optimal classifier, the speaker classifiers are updated $K$ times for each training step in our experiments.

\subsection{Synthesis process}
The synthesizer recovers acoustic features from the concatenation of linguistic representations and speaker label
as $\X' = \S(\H, \y)$, where $\X'=[\x'_1, \dots, \x'_{N_x}]$.
The linguistic and embedded speaker label are repeated to the length of acoustic features and then concatenated as the inputs of
the synthesizer.
The synthesizer architecture
basically follows the decoder in Tacotron model \cite{wang2017tacotron, shen2017natural}. However, it is connected to the recognizer outputs
frame-by-frame rather than utilizing an attention block.
The predicted acoustic features are penalized by the MSE loss as
\begin{equation}
L_{rec} = \frac{1}{N_x} \Sigma_{n=1}^{N_x}{\text{MSE}(\x'_n, \x_n)}.
\end{equation}

Simply applying the MSE criterion often leads to over-smoothed acoustic features.
In order to generate more realistic acoustic features, GAN losses are further incorporated during model fine-tuning.  The recognizer-synthesizer module is used as the generator, i.e. $\X'= \S(\R(\X), \y)$. Speaker-dependent $\D$ is adopted to classify the natural or generated acoustic features for each speaker.
The discriminators are based on 4-layer 1D-CNN followed by a mean pooling layer.
Wasserstein GAN with gradient penalty (WGAN-GP) \cite{pmlr-v70-arjovsky17a, gulrajani2017} is chosen as the objective function in order to stabilize
the training process of GAN. The discriminators are trained with the loss as
\begin{equation}
L_{dis} = \D(\X') - \D(\X) + w_{gp} *  (\parallel \nabla_{\hat{\X}}\D(\hat{\X}) \parallel_2 - 1)^2,
\end{equation}
where $w_{gp}$ represents the weighting factor of GP loss, and $\hat{\X}$ represents randomly sampled features by interpolating between $\X$ and $\X'$. The generator is trained with an adversarial loss
\begin{equation}
L_{gan} = -\D(\X').
\end{equation}

\subsection{Training strategy}
The training process of our proposed model includes pre-training on an external multi-speaker dataset and fine-tuning on
the pair of source-target speakers.
Such design aims to transfer the knowledge learned from large multi-speaker dataset to
one pair of speakers.
It is expected to increase the model's generalization ability especially when the training data
of desired pair is insufficient.
Despite that this paper concentrates on the conversion between a pair of two-speakers, our method can be readily extended to multiple speakers
for many-to-many VC.

In summary, four kinds of losses are imposed during pre-training. They are the phoneme classification loss $L_{p}$, speaker classification losses $L_{s}$ and $L_{s'}$, adversarial losses
 $ L_{adv}$ and $L_{adv'}$, and the reconstruction loss $L_{rec}$. $L_{adv}$ and $L_{adv'}$ are scaled by
$w_{adv}$ and $w_{adv'}$ respectively. Then losses are added
together for training the model.
During fine-tuning, two additional speaker embeddings are initialized randomly while the rest parameters are loaded
from the pre-trained model. In addition to the losses applied during pre-training, GAN losses $L_{dis}$ and $L_{gan}$ are further
adopted. Here, $L_{gan}$ is first scaled by $w_{gan}$ then added to the total loss.
\begin{table}
\renewcommand\arraystretch{1.3}
\caption{Details of  model configurations.}
\label{tab:structure}
\centering
\resizebox{0.9\linewidth}{!}{
\begin{tabular}{c | l}
\hline
\hline

\multirow{4}{*}{ $\R$ }  &Conv1D-k5s2c512-BN-ReLU-Dropout(0.2)  $\to$ \\
                         & 1 layer BLSTM, 256 cells each direction $\to$ \\
                         &  Conv1D-k5s2c512-BN-ReLU-Dropout(0.2)   $\to$ \\
                         & 1 layer BLSTM, 256 cells each direction $\to$ $\H$ \\
\cline{1-2}
$\C_p$                   & one layer LSTM, 128 cells with attention \\
\cline{1-2}

\cline{1-2}
\multirow{2}{*}{$\C_s$}  & Conv1D-k5s1c256-BN-LeakyReLU $\times3$ $\to$ \\
                         & FC-99-Softmax \\
\cline{1-2}
\multirow{6}{*}{$\S$}   & \textbf{Prenet}: FC-256-ReLU-Dropout(0.5) $\times2$  \\
                         &\textbf{RNN}:  2 layer LSTM, 512 cells, \\
                         & 2 frames are predicted each RNN step \\
                         & \textbf{Postnet}: Conv1D-k5s1c256-BN-ReLU-Dropout(0.2) $\times5$ $\to$ \\
                         & Conv1D-k5s1c80, with residual connection \\
                         & from the input to output \\

\cline{1-2}
\multirow{2}{*}{$\D$}    & Conv1D-k5s2c256-LeakyReLU $\times3$ $\to$\\
                        & Conv1D-k5s2c1 $\to$ mean pooling \\

\hline
\hline
\multicolumn{2}{p{215pt}}{
``FC'' represents fully connected layer. ``BN'' represents batch normalization.
``Conv1D-k$k$s$s$c$c$'' represents 1-D convolution with kernel size $k$, stride $s$ and channel $c$. ``$\times N$'' represents
repeating the block for $N$ times. Structure of $\S$ follows the decoder in the Tacotron model \cite{wang2017tacotron, shen2017natural}.}
\end{tabular}}
\end{table}

\begin{table}[t]
  \renewcommand\arraystretch{1.3}
\caption{MCDs and $F_0$ RMSEs on test set using training sets of different sizes. Lower is better.
}\label{tab:objective}
  \centering
  \resizebox{0.75\linewidth}{!}{
  \begin{tabular}{l|c|c|c|c}
  \hline
  \hline
  \multirow{4}*{\# of Utt.}&
  \multicolumn{4}{c}{rms-to-slt} \\
  \cline{2-5}
  &\multicolumn{2}{c|}{VCC2018}& \multicolumn{2}{c}{Proposed}\\
  \cline{2-5}
  & MCD& $F_0$ RMSE& MCD& $F_0$ RMSE \\
  &(dB)&       (Hz)&(dB)&       (Hz) \\
  \hline
  100  & 3.420 &\textbf{14.573} & \textbf{3.323} &18.675 \\
  \hline
  200 & 3.411 & \textbf{15.100} &\textbf{3.252} &16.511  \\
  \hline
  300 & 3.399 &\textbf{14.207} &\textbf{3.246} & 17.134 \\
  \hline
  400 & 3.386 &\textbf{14.784} &\textbf{3.246} &17.357 \\
  \hline
  500 & 3.376 &\textbf{15.042} &\textbf{3.213} & 17.055 \\
  \hline
  \hline
  \multirow{4}*{\# of Utt.}&
  \multicolumn{4}{c}{slt-to-rms} \\
  \cline{2-5}
  &\multicolumn{2}{c|}{VCC2018}& \multicolumn{2}{c}{Proposed}\\
  \cline{2-5}
  & MCD& $F_0$ RMSE& MCD& $F_0$ RMSE \\
  &(dB)&       (Hz)&(dB)&       (Hz) \\
  \hline
  100  &\textbf{3.218}  &\textbf{16.226} &3.286 &18.655  \\
  \hline
  200  &\textbf{3.200}  &\textbf{15.956} &3.245 & 17.546  \\
  \hline
  300 & 3.188 &\textbf{15.455} & \textbf{3.175} &17.638 \\
  \hline
  400 & 3.179 &\textbf{15.595} &\textbf{3.173} &17.204\\
  \hline
  500 & 3.171 &\textbf{15.771} &\textbf{3.147} &17.484 \\
  \hline
  \hline
  \end{tabular}
  }

\end{table}

\section{Experiments}
\subsection{Experimental conditions}
One female speaker (slt) and one male speaker (rms) in the CMU ARCTIC dataset\footnote{\url{http://festvox.org/cmu_arctic/index.html}}
were used as the pair of speakers for conversion in our experiments. For each speaker,
the evaluation and test set both contained 66 utterances. The non-parallel training set
for each speaker contained 500 utterances.
Smaller training sets containing 100, 200, 300 and 400 utterances were also constructed by
randomly selecting a subset of the 500 utterances for training.
The multi-speaker VCTK dataset \cite{veaux2017cstr} was utilized for model
pre-training. Altogether 99 speakers were selected from VCTK dataset.
For each speaker, 10 and 20 utterances were used for validation and testing repsectively.
The remaining utterances were used as training samples. The total duration of training samples
was about 30 hours.

For acoustic features, 80-dimensional Mel-spectrograms were extracted every 10 ms and then scaled
to logarithmic domain.
Adam \cite{KingmaB14} optimizer was used with a learning rate of 0.001.
The batch size was 32 and 8 at the pre-training and fine-tuning stage respectively.
The weighting factors of adversarial losses were set as $w_{adv}=100$, $w_{adv'}=5$ and $w_{adv}=1, w_{adv'}=0.1$
during pre-training and fine-tuning respectively. $K$ was set as 2. For the GAN loss, $w_{gp}$ and $w_{gan}$ were set as $10$ and $0.05$
respectively.  After fine-tuning, the accuracy of the speaker classifier on the test sets of slt and rms was 72.2\%. In comparison, it was 100.0 \% without using adversarial losses. And the accuracy of phoneme classifier was 89.4\%.

The details of our model structure are summarized in \tablename~\ref{tab:structure}.
The implementation of WaveNet vocoder
followed our previous work \cite{ljliu2018wav}.
Since this paper focuses on the acoustic models for VC, the
same WaveNet vocoders trained with 500 utterances were used when varying the size of data
for fine-tuning acoustic models.

We compared our proposed method with a Rec-Syn baseline \cite{ljliu2018wav} (i.e., VCC2018)\footnote{
Audio samples of our experiments are available at \texttt{\url{https://jxzhanggg.github.io/advVC/}}.}.
In this method, bottleneck features were extracted by an ASR model trained on about 3000 hours of external speech data
as linguistic descriptions and were used as the inputs of speaker-dependent synthesis models.
This method achieved the best
performance on the non-parallel VC task of Voice Conversion
Challenge 2018.

\subsection{Objective evaluation}

For objective evaluation,
$F_0$ and 25-dimensional MCCs features were extracted by STRAIGHT \cite{Kawahara1999Restructuring}
from the reconstructed waveforms for evaluation. Then, Mel-cepstrum distortions (MCD)
and root mean square error of $F_0$ ($F_0$ RMSE) on test set were
reported in Table~\ref{tab:objective}.

Compared with the VCC2018 baseline, our proposed method achieved lower MCD except in
slt-to-rms conversion given 100 and 200 training utterances.
However, for $F_0$ RMSE metric, the VCC2018 achieved better results compared to the proposed method.
This results indicated the potential of further improving $F_0$ prediction in our proposed method.
We should notice that VCC2018 method exploited a large amount of data (i.e., 3000 h) for training the ASR model.
On the other hand, the proposed method was pre-trained on much smaller VCTK dataset (i.e., 30 h).

\begin{table}
\renewcommand\arraystretch{1.3}
\caption{MCDs and $F_0$ RMSEs in ablation studies of proposed method. Lower is better.}
\label{tab:ablation}
\centering
\resizebox{0.75\linewidth}{!}{
\begin{tabular}{c|c|c|c|c}
\hline
\hline

\multirow{3}{*}{ Methods }  & \multicolumn{2}{c|}{rms-to-slt} & \multicolumn{2}{c}{slt-to-rms} \\
\cline{2-5}
                            & MCD & $F_0$ RMSE  & MCD & $F_0$ RMSE \\
                            & (dB) & (Hz)& (dB) & (Hz) \\
\hline
Proposed &\textbf{3.213} &17.055 &\textbf{3.147} &\textbf{17.484} \\
\hline
\emph{-adv} &3.967 &29.140 & 3.683 & 22.929 \\
\hline
\emph{-phone} &3.781 &22.232 & 3.753 & 20.038 \\
\hline
\emph{-pretrain} &4.228 &27.177 & 3.911 & 44.790 \\
\hline
\emph{-joint} &3.267 &17.223 & 3.214 & 17.550 \\
\hline
\emph{-tunerec} &3.444 &\textbf{16.905} & 3.411& 18.968 \\
\hline
\emph{-all} &4.287 & 24.443 & 3.900& 35.969 \\
\hline
\hline
\end{tabular}}
\end{table}

In order to analyze the effects of various strategies used in our model, ablation studies were further conducted.
For investigating the effects of speaker adversarial training, we removed the losses of $L_{adv}$ and $L_{adv'}$
(i.e., \emph{``-adv''}).
For investigating the effects of phoneme classification, the loss $L_{p}$ was removed (i.e., \emph{``-phone''}).
For investigating the effects of pre-training, the model was initialized randomly before fine-tuning (i.e., \emph{``-pretrain''}).
For
investigating the effects joint optimization, the recognizer and synthesizer were trained separately (i.e., \emph{``-joint''}).
An experiment was also conducted that
fixed the recognizer and only adapted the synthesizer
  on the target speaker during fine-tuning (i.e., \emph{``-tunerec''}).
In analogy to the VCC2018 baseline,
a conventional Rec-Syn model was built (i.e., \emph{``-all''})
using the same training data and model structure as those of our proposed method.
In this method, the recognizor was first trained with the phoneme classification loss for
extracting linguistic features. Then,
the synthesizer was pretrained and finetuned on the target speaker.

\tablename~\ref{tab:ablation} summarizes the results of ablation studies.
From the table, we can see that
performance of the proposed method degraded without using
either the speaker adversarial loss or the phoneme classification loss.
When listening to the converted samples for further examination,
it's found that the voice converted by \emph{``-adv''} method suffered from low similarity
while those converted by \emph{``-phone''} method had low intelligibility.
For the \emph{``-pretrain''} method, objective errors increased drastically.
And the converted voice was hardly intelligible.
Objective errors slightly rose when using the \emph{``-joint''} method.
It indicated that training the recognizer and the synthesizer separately leaded to sub-optimal solutions.
For the \emph{``-tunerec''} method, the spectral distortion increased and
the $F_0$ error was close to the proposed method.
These results indicated fine-tuning the whole model on
both source and target data improved the performance of the model.
From the last row of the table, we can see the improvement of our proposed method over conventional Rec-Syn method
was significant .

\begin{figure}[t]
    \centering
    \centerline{\includegraphics[width=0.80\linewidth]{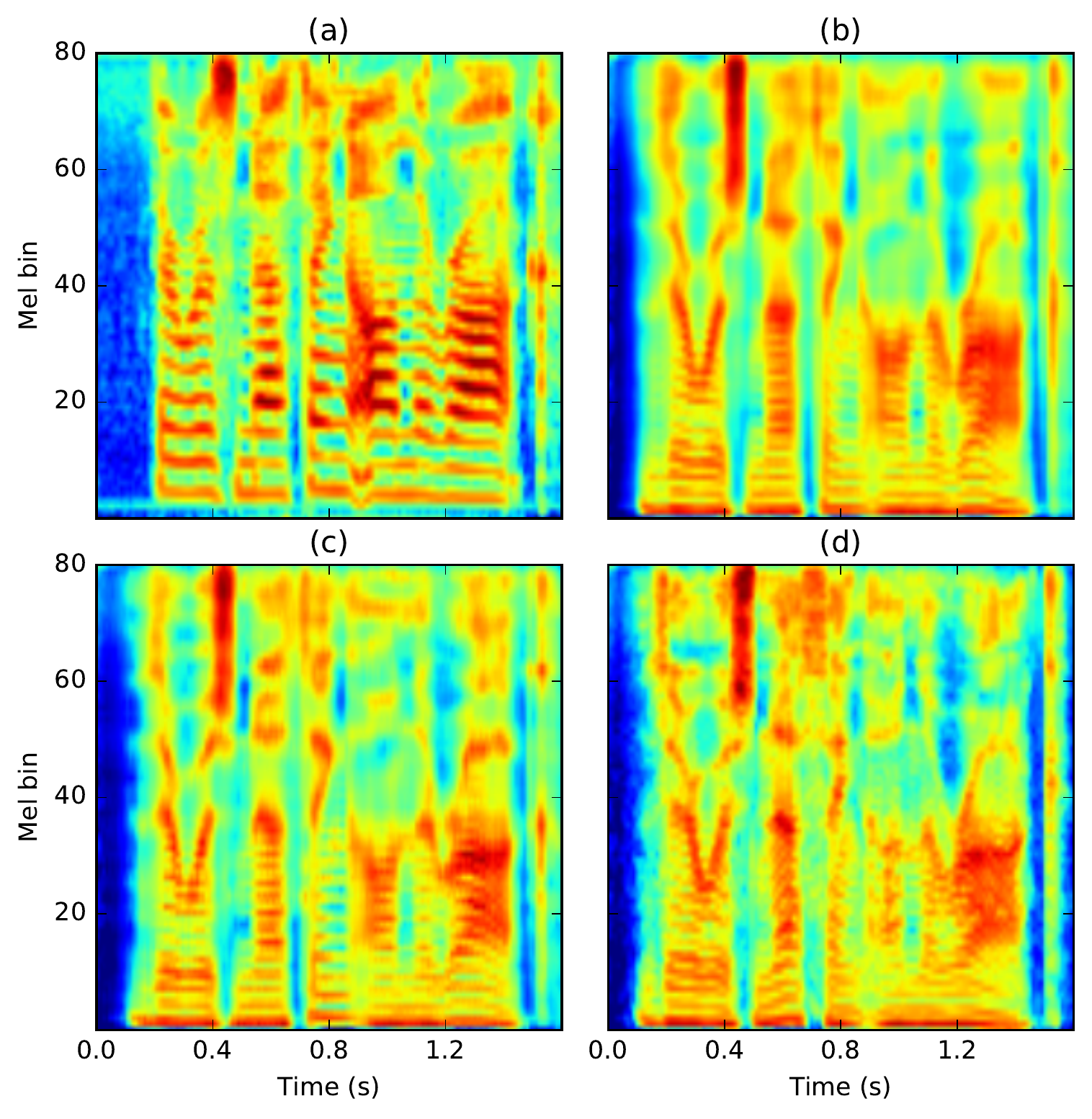}}
    \caption{Mel-spectrograms of (a) a source utterance, (b) the voice converted by our proposed method
    without GAN loss, (c) the voice converted by our proposed method and (d) the target utterance.}
\label{fig:spc}
\end{figure}

\figurename~\ref{fig:spc} (a) and (c) show the Mel-spectrograms of one source utterance and its converted voice using our proposed method respectively. The converted Mel-spectrogram is similar to that of natural reference in \figurename~\ref{fig:spc} (d).
Comparing the Mel-spectrogram converted by our proposed method to that converted by the method without GAN loss in \figurename~\ref{fig:spc} (b), we can see that the GAN loss helped to alleviate the over-smoothing problem and to enhance the format structures.

\subsection{Subjective evaluation}

The ``\emph{-all}'' method in previous ablation study, the VCC2018 baseline and the proposed method were compared in subjective evaluations.
For each experiment, at least thirteen listeners were involved.
Samples were presented to them using headphones in random order.
They were asked to give a 5-scale opinion score (5: excellent, 4: good, 3: fair, 2: poor, 1: bad) on both similarity and naturalness for each converted utterance.
20 utterances were selected randomly from the test set and two conversion directions (i.e., slt-to-rms and rms-to-slt) were evaluated for each method.

\begin{table}
\renewcommand\arraystretch{1.3}
  \caption{Mean opinion scores with 95\% confidence intervals of different methods on test set. Higher is better.}
  \label{tab:subjective}
  \centering
  \resizebox{0.9\linewidth}{!}{
  \begin{tabular}{c c|c|c|c}
  \hline
  \hline
 \multicolumn{2}{c|}{\# of Utt.}  &\emph{-all} & VCC2018 &Proposed \\
  \hline
  \multirow{2}{*}{100} & Nat. &1.514 $\pm$ 0.091 &\textbf{3.714} $\pm$ 0.130 &3.628 $\pm$ 0.119 \\
  \cline{2-5}
                       & Sim. &1.471 $\pm$ 0.086 &3.764 $\pm$ 0.153 &\textbf{3.850} $\pm$ 0.134 \\
  \hline
    \multirow{2}{*}{500} & Nat. &1.493 $\pm$ 0.093 &3.636 $\pm$ 0.132 &\textbf{3.950} $\pm$ 0.101 \\
  \cline{2-5}
                       & Sim. &1.457 $\pm$ 0.088 &3.685 $\pm$ 0.154 &\textbf{4.129} $\pm$ 0.120 \\
  \hline
  \hline
  \end{tabular}}
\end{table}

From \tablename~\ref{tab:subjective}, we can see that the
proposed method improved the
naturalness and similarity
of the \emph{``-all''}  method with a large margin. It indicated
that our proposed method exploited training data more efficiently with adversarial learning.
Our method outperformed the VCC2018 baseline given 500 training utterances of both speakers for fine-tuning, in terms of both naturalness
and similarity.
In the condition of using 100 training utterances,
our method achieved higher similarity while lower naturalness
than the VCC2018 method.
Despite that our method could obtain better disentangled representations,
the VCC2018 baseline learned more fine-grained linguistic descriptions
by training on large external corpus. This is especially favorable when the training data of the conversion pair is scarce.

\section{Conclusions}
In this paper, a method for non-parallel voice conversion is proposed.
Our model is based on the recognition-synthesis framework and a speaker classifier module
is introduced for speaker adversarial learning.  We also
incorporate GAN losses for boosting the quality of converted
voice.
The model is first pre-trained on a multi-speaker dataset then
 fine-tuned on the desired conversion pair.
Both objective and subjective evaluations proved the
effectiveness of our method. Our future work will try to further improve the performance of our method by pre-training on larger datasets.

\bibliographystyle{IEEEtran}

\bibliography{mybib}

\end{document}